# Statistical and Clinical Aspects of Hospital Outcomes Profiling

**Sharon-Lise T. Normand and David M. Shahian**

*Abstract.* Hospital profiling involves a comparison of a health care provider's structure, processes of care, or outcomes to a standard, often in the form of a *report card*. Given the ubiquity of report cards and similar consumer ratings in contemporary American culture, it is notable that these are a relatively recent phenomenon in health care. Prior to the 1986 release of Medicare hospital outcome data, little such information was publicly available. We review the historical evolution of hospital profiling with special emphasis on outcomes; present a detailed history of cardiac surgery report cards, the paradigm for modern provider profiling; discuss the potential unintended negative consequences of public report cards; and describe various statistical methodologies for quantifying the relative performance of cardiac surgery programs. Outstanding statistical issues are also described.

## 1. INTRODUCTION

Profiling involves a comparison of a health care provider's structure, processes of care, or outcomes to a normative or community standard (Gatsonis, 1998). The results of such profiling are typically presented in the form of a report card, whose purpose is to quantify quality of care (Normand, 2005). This quality triad was first conceptualized by Avedis Donabedian (Donabedian, 1980), a Distinguished University Professor of Public Health at the University of Michigan whose work was devoted to the study of health care quality. Structural measures include, for example, nursing ratios, the presence of residency programs, availability of advanced technology and procedural volume. Such measures are straightforward and relatively easy to define, but their precise relationship to actual health outcomes is often difficult to quantify. For example, volume has been shown to be a good quality surrogate for many surgical procedures. However, for coronary artery bypass grafting (CABG) surgery, an operation that creates a new route around a blocked portion of an artery in order to increase blood supply to the heart muscle, this relationship is relatively weak (Peterson et al., 2004; Shroyer et al., 1996; Shahian and Normand, 2003).

Process measures refer to what providers do to and for patients. These include documented adherence to established best practices, such as the use of peri-operative beta blockade to reduce myocardial ischemia, or time from hospital arrival to angioplasty for acute myocardial infarction in order to limit irreversible heart muscle injury. Process measures may be available for many conditions where outcome measures do not exist or have limited applicability due to sample size issues or infrequent endpoints. Another attractive feature is that they are transparent and actionable, giving providers immediate guidance as to where to focus improvement efforts. Process measures also have the advantage of not requiring direct risk adjustment, but legitimate exclusions from the denominator, such as contraindications to the recommended practice, must be recognized and documented. With the increasing availability of evidence-based practice guidelines, there has been a corresponding increase in the number of candidate measures for process-based quality improvement. However, there is some concern that excessive emphasis on achieving compliance with process measures stimulated by public reporting or pay-for-performance initiatives might lead to unnecessary screening procedures or treatments, or that they might conflict with a physician's best judgment or patient preference (Werner and Asch, 2005). Furthermore, process measures focus on only a limited segment of the overall care provided for a particular medical condition; they may be applicable to only a small percentage of patients with a given condition; and they explain a relatively small percentage of the variability in outcomes, which is the real interest of patients (Krumholz et al., 2007; Werner and Bradlow, 2006; Bradley et al., 2006; Rathore et al., 2003).

*Sharon-Lise T. Normand is Professor, Department of Health Care Policy, Harvard Medical School, 180 Longwood Avenue, Boston, Massachusetts 02115, USA e-mail: sharon@hcp.med.harvard.edu. David M. Shahian is Professor, Tufts University School of Medicine, Boston, Massachusetts 02111, USA.*

Since the publication of several Institute of Medicine reports on quality, transparency and patient safety (Institute of Medicine, 2000, 2001), attention has increasingly been focused on the objective measurement of health care outcomes. Outcomes measures refer to responses that characterize the patient's health status and include, for example, perioperative mortality, morbidity and functional status. Coronary artery bypass grafting surgery is one of the few procedures performed with sufficient frequency to justify statistical assessment of outcomes to quantify provider performance (Dimick, Welch and Birkmeyer, 2004). It is the outcomes component of Donabedian's triad that requires the most sophisticated statistical approach and this is the primary motivation for this article.

As described by Birkmeyer, Dimick and Birkmeyer (2004), the choice of structural, process or outcome measures to assess quality is dependent on both the overall procedural frequency and the potential for serious adverse consequences. The focus of this article on public profiling of hospital outcomes is not intended to diminish the importance of either confidential continuous quality improvement (CQI) activities or profiling based on process or structural measures. For example, in cardiac surgery, substantial improvements in overall regional CABG surgery outcomes and a marked reduction in interprovider variability have been achieved through a confidential CQI approach, which is a completely different paradigm than public report cards (O'Connor et al., 1996). Furthermore, the Society of Thoracic Surgeons Quality Measurement Taskforce has developed a multidimensional composite measure of cardiac surgery quality that encompasses each of the components of Donabedian's triad. Notwithstanding these caveats, in the current era of health care performance transparency, public accountability for outcomes is unquestionably of major importance to payors, regulators and consumers.

## 2. CLINICAL CONSIDERATIONS

### 2.1 Background

From a historical standpoint, measurement of health care outcomes has a long but rather sparse history. The basic concept of comparative profiling has been understood for over 150 years. Florence Nightingale, born in Italy in 1820 of English parents, obtained an early education that was notable for its unusually strong emphasis in mathematics. She felt called to the profession of nursing and ultimately led a group of nurses to Scutari during the Crimean War (Cohen, 1984; Iezzoni, 1997a, 2003; Stinnett et al., 1990; Spiegelhalter, 1999). Conditions in British hospitals were deplorable, and Nightingale was able to accumulate extensive data on the causes of death among British soldiers, subsequently displaying these outcome data in "coxcomb" or polar area charts. These demonstrated that diseases related to poor sanitary conditions in such hospitals killed many times more soldiers than war wounds. Upon her return to England, she continued her analytical studies by comparing mortality statistics among London hospitals (Iezzoni, 1997a, 2003). Significant disparities in outcomes were noted, many of which were felt to be due to overcrowding and generally unsanitary conditions. In various articles, she also noted the importance of accounting for patient status on admission, presaging current efforts to account for case mix through risk adjustment. Finally, she also noted that some hospitals intentionally discharged terminal patients, only to have their death occur in another institution. This practice unfairly biased institutional performance comparisons, and it is one form of outcomes gaming (Shahian et al., 2001). For these and numerous other reasons, contemporary researchers have proposed that in-hospital mortality is an inadequate measure of performance and that 30-day follow-up or even longer should be considered the norm (Osswald et al., 1999; Shahian et al., 2001). Because of her contributions to applied statistics and epidemiology, with particular emphasis on health care improvement, Nightingale was made a Fellow of the Royal Statistical Society and an Honorary Member of the American Statistical Association, both rare honors for a woman of her era (Cohen, 1984; Iezzoni, 1997a, 2003; Spiegelhalter, 1999; Stinnett et al., 1990).

Unquestionably, the most significant early pioneer in the area of outcomes measurement, analysis and public report cards was Dr. Ernest Amory Codman, a turn-of-the-century Boston Brahmin who graduated from Harvard College and Harvard Medical School (Iezzoni, 1997a, 2003; Passaro and Organ, 1999; Spiegelhalter, 1999; Mallon, 2000). Like Florence Nightingale, he had always displayed a passion for objective quantitative data, and he maintained a log of shells expended to birds shot when hunting (Passaro and Organ, 1999). While Nightingale's approach was more epidemiologic, Codman focused more on surgical audit of individual cases, classification of surgical errors and comparison of outcomes (Spiegelhalter, 1999). He developed the first anesthesia chart as well as a unique system of classifying surgical errors. Although on the surgical staff at the Massachusetts General Hospital, he started his own hospital on Pinckney Street in Boston in 1911 and maintained meticulous records of both short- and long-term outcomes. His incessant plea for hospitals to maintain accurate records of patient outcomes, to accept responsibility for these outcomes and to publicize them were not well received. Together with his suggestion that the seniority system for selecting surgical leadership should be eliminated, this ultimately led to his resignation from the staff at the Massachusetts General Hospital and his alienation from much of the Boston medical community. He nonetheless persevered and was instrumental in the development of the American College of Surgeons. He served as Chair of its Committee on Hospital Standardization, a forerunner of the Joint Commission on Accreditation of Healthcare Organizations (Passaro and Organ, 1999; Mallon, 2000).

It is to giants such as Florence Nightingale and Ernest Amory Codman, considered by many to be iconoclasts in their own day, that our modern approach to outcomes analysis owes incalculable debt.

## 2.2 Cardiac Surgery Profiling

The modern era of publicly profiling institutional performance began not in health care but in education, most notably in Great Britain (Aitkin and Longford, 1986; Goldstein and Spiegelhalter, 1996; Jaeger, 1989). Studies of school effectiveness anticipated many of the statistical issues and controversies currently being debated in health care. Profiling initiatives in the latter area arguably began in 1986 with the first annual report of hospital-level data for 17 broad diagnostic and procedure groups, released by the Health Care Financing Administration (HCFA), now known as CMS, the Center for Medicare and Medicaid Services. CMS is the U.S. federal agency that administers the Medicare program and works with states to administer Medicaid and the State Children's Health Insurance Program. Hospitals in the 1986 report with higher than expected mortality rates were classified as having potential quality problems. These reports were widely criticized for their failure to adequately account for patient severity, not to mention numerous other anomalies such as high mortality rates attributed to a hospice (Berwick and Wald, 1990; Iezzoni, 1997a; Kassirer, 1994). The public release of hospital report cards was suspended in 1994 (Kassirer, 1994). Nonetheless, certain specialties recognized that, however flawed, this report signaled the beginning of a new era of increased accountability in health care that could not be ignored. Cardiac surgery was the first and most prominent of these specialties, and it has unquestionably become the paradigm for modern health care outcomes measurement. Coronary artery bypass grafting (CABG) surgery is the focus of many profiling efforts. It is the most commonly performed major complex surgical procedure, it is costly, and it has well-defined endpoints including serious complications and death (Shahian et al., 2001).

Soon after the publication of the HCFA report card, the Society of Thoracic Surgeons established an Ad Hoc Committee on Risk Factors for CABG, and the Society also began work on development of what ultimately became the Society of Thoracic Surgeons National Cardiac Database (STS NCD), which was released to its membership in 1990. During the same time period, other seminal studies of risk-adjusted cardiac surgery outcomes demonstrated unexpected and significant variability in outcomes not accounted for by case mix. In an analysis of 7596 New York State patients who had undergone open heart surgery during the first six months of 1989, Hannan et al. (1990) noted that unadjusted institutional mortality rates varied from 2.2% to 14.3%, which was a much greater range than expected. The Northern New England Cardiovascular Study Group found that among five centers and 18 cardiothoracic surgeons who performed CABG surgery in Maine, New Hampshire and Vermont between 1987 and 1989, the unadjusted in-hospital mortality varied from 3.1% to 6.3% among hospitals, and a significant difference was noted even after risk adjustment (O'Connor et al., 1991). Williams, Nash and Goldfarb (1991) studied CABG results from 1985 to 1987 at five Philadelphia teaching hospitals and found a twofold variation in mortality for DRG 106 (coronary artery bypass surgery without coronary catheterization) that was not accounted for by the limited risk adjustment that was available.

These early findings stimulated the further application of statistical risk models to assist in (a) the identification of pre-operative factors affecting CABG surgery outcomes; (b) patient counseling; (c) procedure selection; (d) outcomes assessment and profiling; and (e) continuous quality improvement activities (Shahian et al., 2001, 2004). Subsequently, similar models have been developed for post-cardiac surgery complication rates and also for valvular heart surgery, congenital heart surgery and general thoracic surgery (Shahian et al., 2004). In addition to the STS NCD, other excellent risk models continue to be employed by the New York Cardiac Surgery Reporting System (CSRS), the Northern New England Cardiovascular Disease Study Group (NNE), the Veterans Affairs Administration and a European consortium (Shahian et al., 2001, 2004).

By far the most controversial use of risk models has been for the determination and comparison of risk-adjusted mortality rates for hospitals and individual surgeons. Typically, the probabilities of death for all of a provider's patients are estimated from logistic regression, aggregated, and compared with the observed number of deaths, usually by means of a ratio of the observed number to the expected number. This may then be multiplied by the overall unadjusted mortality for a state or region, yielding a so-called risk-standardized mortality rate. Numerous statistical concerns have been expressed regarding this approach, including the inaccuracy of estimates from low-volume providers with small sample sizes, clustering (nonindependence) of patients among providers, and multiple comparisons (Thomas,

Longford and Rolph, 1994; Goldstein and Spiegelhalter, 1996; Christiansen and Morris, 1997; Normand, Glickman and Gastonis, 1997; Shahian et al., 2001, 2004).

Few published studies have correlated risk-standardized mortality rates with objective or subjective expert assessment of quality. In hospital site visits, Daley et al. (1997) found differences in processes and delivery of surgical care in Veterans Affairs Medical Centers that correlated with and corroborated statistical measures of high hospital risk-adjusted mortality and morbidity. In a recent study of hospital process measures and in-hospital mortality among patients with acute coronary syndromes, Peterson et al. (2006) found significant associations between use of needed therapies and mortality.

2.2.1 *Data quality.* In any profiling initiative, the quality of the data is more important than choice of statistical models. Clear and concise definitions for data elements are exceedingly important, especially for those variables that are most highly predictive of mortality. Coding accuracy may significantly affect risk-adjusted outcome results. A prospectively maintained clinical database containing core clinical variables is the best data source for profiling (Shahian et al., 2001, 2004; Krumholz et al., 2006a).

Administrative claims data, consisting of demographic, diagnosis and procedural codes are derived primarily from insurance claims. These data are readily available for millions of patients, but they are not collected with the primary goal of assessing risk-adjusted patient outcomes. Cases may be missed or misclassified, and important but nonreimbursable diagnoses may be excluded (Krumholz et al., 2006a; Iezzoni, 1997a,b, 2003; Shahian et al., 2001, 2004). In a recent study by Mack et al. (2005), CABG cases at one Texas hospital were analyzed to determine whether there was agreement between the results from an audited clinical database (STS NCD) and federal and state administrative databases. There were significant disparities in both the volumes of cases and the unadjusted mortality rates with administrative data significantly overstating the latter. Similar case misclassification errors were observed in Massachusetts when the results from a carefully audited registry using clinical data (STS NCD) were compared with those from a state administrative database (Shahian et al., 2007).

Similarly, separation of pre-operative co-morbidities (case-mix adjustors) from post-operative complications is problematic when using administrative data sources (Iezzoni, 1997a,b, 2003). Such misclassification may lead to a provider in essence receiving credit for operating upon a patient with a serious pre-existing condition, when in fact that condition was a major complication of care. For example, poor surgical care may lead to pulmonary complications such as pneumonia; inclusion of pneumonia in the statistical model would actually "adjust away" the outcome. Conversely, misclassifying a true pre-operative risk factor as a complication fails to account for case-mix acuity, thus disadvantaging a provider's risk-adjusted outcomes. This critical deficiency of administrative databases has led to the development of numerous computerized algorithms for correctly classifying secondary diagnoses, none of which is foolproof and all of which are inferior to the definitive classification available in a clinical database. When only administrative data are available, date stamping or "present at admission" indicators are the best compromise solution for correctly classifying secondary diagnoses. The Office of Statewide Health Planning and Development in California, for example, creates inpatient hospital discharge data files that separate conditions present on admission from those present on discharge.

A comprehensive analysis of the impact of comorbidity undercoding has recently been conducted by Austin et al. (2005) using Monte Carlo simulation techniques. Although the assignment of outlier status was relatively robust to undercoding of severity and co-morbidities, miscoding of very influential predictors, such as shock or renal failure, could lead to hospital misclassification.

To assess the practical impact of using administrative data for profiling, Hannan and associates compared New York CABG results determined from their dedicated clinical database (CSRS) with those derived from the New York administrative database (Hannan et al., 1992) and the federal Medicare administrative database (Hannan et al., 1997a). Not surprisingly, models based upon clinical data provided superior discrimination and accuracy in explaining variations in patient mortality. Models derived from administrative data had significantly improved performance when a few critical clinical variables were added, such as ejection fraction, re-operation or left main coronary artery disease. Studies from the Cardiac Care Network of Ontario (Tu, Sykora and Naylor, 1997), the Cooperative CABG Database Project (Jones et al., 1996) and the STS NCD (Shahian et al., 2004) suggest that a few critical core variables provide much of the important predictive information in any cardiac surgery database.

Notwithstanding these legitimate concerns, for certain conditions administrative data may be sufficient to report on some outcomes. For example, in studies of acute MI and heart failure, Krumholz et al. (2006b,c) found that while models based on medical record data had better discrimination between survival and mortality for individual patients than administrative data, there were not many differences between the hospital-specific standardized risk-adjusted rates using these two data sources.

Inclusion of demographic and socio-economic status (SES) variables as adjustors in the model raises concerns similar to those related to inclusion of complications. Disparities in outcomes among racial/ ethnic groups may be due to system-level factors, such as financing, structure of care and cultural-linguistic barriers; patient-level factors, such as preferences and biological differences; or physician or provider factors, such as bias (Institute of Medicine, 2003). Inclusion of race/ethnicity would only make sense in a profiling context if there were biological differences in survival or patient preferences that impacted survival among different racial/ethnic groups. In CABG surgery, for example, because women are more likely to have smaller vessels that are technically more difficult to bypass, sex is included in the mortality model. However, adjusting for race/ethnicity may in fact unfairly mask those institutions that have poor systems of care such as hospitals that lack translators or hospitals that provide suboptimal care to patients with fewer financial resources. SES variables may be used to help understand differences in quality, such as access barriers, but they should not be used to quantify deficiencies in hospital performance.

### 2.2.2 *Audit and validation.*
Comprehensive data audit and validation are critical to any profiling effort, and their absence is a significant theoretical deficiency of many voluntary initiatives and of registries based on administrative data. We illustrate the importance of these measures with a description of the processes employed in the implementation of the first Massachusetts public report card for cardiac surgery (Shahian, Torchiana and Normand, 2005). Like New York, New Jersey and Pennsylvania (Shahian et al., 2001), Massachusetts has a mandated surveillance program for invasive cardiac services that includes public reporting of hospital mortality (Shahian, Torchiana and Normand, 2005). The first cardiac surgery report card in Massachusetts was based upon data from 2002, during which time frame there were 4603 isolated CABG patients available for analysis (Table 1), distributed among 11 established cardiac surgery programs and two new programs (Hospitals 5 and 10). The first cardiac surgery was performed at Hospital 10 during April 2002 and at Hospital 5 that August. These outcome data were collected, cleaned, audited, validated and analyzed at a central data coordinating center (www.massdac.org). Data submissions were validated using both state administrative and vital statistics databases. One hundred and fourteen data quality reports were given to hospitals with the expectation that problematic data would be corrected or appropriate documentation provided. An audit of 500 cases including all deaths was performed by the local Quality Improvement Organization. A second audit of 724 charts was performed by an expert Adjudication Committee consisting of three senior Massachusetts cardiac surgeons. This focused primarily on data elements that were particularly important in the risk model (e.g., urgent and emergent status, cardiogenic shock, etc.) as well as cases coded as *CABG plus other*, a category potentially used to hide mortalities. Additional documentation that was typically requested by the Adjudication Committee included history and physical examinations, progress notes, operative notes, ICU flow charts and discharge notes. Eight hundred and thirty-five changes were made by the committee, each of which required unanimous agreement.

## 2.3 Report Card Controversies

Aside from issues relating to implementation (type of database, inclusion of critical core variables, audit and validation, and the selection and development of statistical models), there are numerous philosophical and practical concerns regarding both the efficacy and potential unintended negative consequences of report cards (Shahian et al., 2001, 2004; Werner and Asch, 2005). Report cards provide transparency and public accountability, which are perhaps sufficient justification for their existence (Shahian et al., 2001). However, the market-based assumption that consumers will seek to choose the best providers and that providers will respond by improving their quality is as yet unproven (Shahian et al., 2001; Werner

Table 1
*30-day mortality in 13 nongovernmental hospitals following isolated CABG surgery, Massachusetts, USA*

| Cardiac surgery program (1) | Number of cases (2) | Number (%) of deaths (3) | Expected mortality % (4) |
|---|---|---|---|
| 1 | 508 | 11 (2.17) | 2.01 |
| 2 | 454 | 11 (2.42) | 2.58 |
| 3 | 381 | 15 (3.94) | 2.94 |
| 4 | 623 | 11 (1.77) | 2.30 |
| 5 | 26 | 0 | 1.10 |
| 6 | 393 | 7 (1.78) | 2.15 |
| 7 | 718 | 18 (2.51) | 2.20 |
| 8 | 149 | 1 (0.67) | 1.45 |
| 9 | 80 | 0 | 0.87 |
| 10 | 296 | 5 (1.69) | 1.99 |
| 11 | 191 | 3 (1.57) | 1.71 |
| 12 | 365 | 4 (1.10) | 1.87 |
| 13 | 419 | 15 (3.58) | 1.91 |
| All | 4603 | 101 (2.19) | |

Data correspond to surgeries performed in adults during the period January 1, 2002 through December 31, 2002.
(2): Number of admissions in which the first surgery was an isolated CABG surgery.
(3): Number (percent) of observed 30-day mortalities.
(4): Rates expected using estimates of the association between risk factors and mortality, ignoring hospital effects [Section 3.3, (6)].

and Asch, 2005). Furthermore, public report cards may be an incentive for certain behaviors that actually decrease overall health care quality (Dranove et al., 2003).

2.3.1 *Impact on mortality.* Although there was a substantial decline in New York State cardiac surgery mortality that coincided with the introduction of public report cards, it is unclear whether publication of results was the primary mechanism (Shahian et al., 2001). Collecting and analyzing their own data forces hospitals to confront inferior results, and to institute changes to procedures and staff long before such results could ever be disseminated to the public. Furthermore, the decline in New York CABG mortality occurred during a period when these same rates were falling nationally. Ghali et al. (1997) compared the mortality decline in New York with the results from a comparable time frame in northern New England and Massachusetts and found little difference in the magnitude of change. Northern New England had a voluntary, confidential approach to continuous quality improvement (CQI) with no public reporting, and Massachusetts had strong aca-demic and clinical centers but no public report cards. In a study of Medicare CABG patients, Peterson et al. (1998) reviewed results for all states and regions from 1987 to 1992. The only geographic region with a level of outcomes improvement and low absolute CABG mortality level comparable to New York's was northern New England, which as noted has a totally confidential CQI approach. When the first official Massachusetts report card was published in 2004, employing highly audited clinical data and a sophisticated statistical model, it was observed that the 2002 unadjusted mortality rate for CABG was 2.19%, arguably one of the lowest overall state CABG mortality rates ever reported. This was in a state which had never before had a public report card. Such observations suggest that report cards, although useful for public accountability, are only one of many motivating factors for high quality, and perhaps not essential.

2.3.2 *High-risk case avoidance.* Analysis of New York clinical CABG data by Hannan et al. (1997b) suggests that modern risk-adjustment algorithms do in fact adequately protect providers who undertake the care of high-risk patients. However, despite the availability of such sophisticated CABG risk models, the threat of public disclosure of results has inevitably resulted in more selective acceptance of patients by cardiac surgeons. Omoigui et al. (1996) noted that after the introduction of report cards in New York State, more New York patients with high-risk characteristics were sent to the Cleveland Clinic for cardiac surgery than had been referred during the pre-report period. Furthermore, these New York State patients at the Cleveland Clinic had the highest expected mortality of any referral group there, and referrals from New York to Cleveland increased during the post-report card period in contrast to all other states. Subsequent studies by Chassin, Hannan and DeBuono (1996) and Peterson et al. (1998) have challenged this out-migration hypothesis. However, it is undeniable that many surgeons perceive that accepting such high-risk patients may jeopardize their reputations and referrals. In a study of Pennsylvania cardiac surgeons by Schneider and Epstein (1996), 63% reported that they were less willing to operate on severely ill patients subsequent to the introduction of report cards. Furthermore, 59% of cardiologists reported increased difficulty finding surgeons willing to accept such high-risk patients.

Burack et al. (1999) found that high-risk CABG patients in New York, whose results would be publicly reported, were more likely to be refused surgery than were similar high-risk patients with aortic dissection, another type of cardiac surgery for which results are not reported. In that study, 62% of cardiac surgeons reported that they had refused to operate on at least one high-risk patient during the preceding year because of the fear of public reporting. Numerous solutions to this problem have been recommended, such as the exclusion of high-risk patients from reporting, compiling data on cardiac patients from the time of initial referral in order to track inappropriate denials of surgical care, and the collection of other quality indicators in addition to mortality (e.g., morbidity, quality of life and functional improvement) (Shahian et al., 2001). There is no question that focusing on public reporting of mortality will result in some biasing of patient selection and may deny surgical intervention to the very group of high-risk patients who might benefit most (Jones, 1989). Furthermore, patients denied appropriate CABG may be subjected to less effective and cumulatively more costly therapies, leading to both higher societal costs and overall population mortality rates (Jones, 1989; Dranove et al., 2003).

2.3.3 *Gaming.* Diagnostic Related Groups (DRG)-based reimbursement strategies led to the development of "DRG creep," in which institutional coding practices changed in order to maximize hospital reimbursement. Similarly, when faced with the prospect of public outcomes reports that may impact licensure, referrals and pay-for-performance reimbursement, surgeons and institutions may attempt to "game" the outcome reporting system. For example, by inappropriately coding pre-operative co-morbidities, especially those like emergency status or cardiogenic shock that are highly predictive of operative mortality, a provider's expected mortality is increased, and their O/E ratio and risk-adjusted mortality rate decrease (Greene and Wintfeld, 1995; Parsonnet, 1995; Shahian et al., 2001). Careful audit is essential to detecting and discouraging such practices, particularly when the frequency of co-morbidities for a particular institution is out of the usual range. Change of operative class is another form of gaming. If only isolated CABG procedures are publicly reported, a surgeon who anticipates a bad patient outcome may add a relatively trivial additional component to the operation, such as closure of a patent foramen ovale. This would remove the procedure from the isolated CABG category which is publicly reported, and shift it into the *CABG plus other* category which is unreported. Careful audit of all *CABG plus other* cases is essential to detecting such practices, and it is also critical to define prospectively what types of cases truly justify an *other* designation. For example, in Massachusetts it was decided that CABG plus closure of a patent foramen ovale would be coded as an isolated CABG. It is a relatively trivial procedure that does not really change the expected mortality, but it could theoretically be applied to a significant percentage of the CABG population.

For purposes of outcome reporting, it is also important to follow patients for at least 30 days following major surgery, if not longer (Osswald et al., 1999). With modern technology, many patients who are very seriously ill after CABG may be kept alive for weeks or months, only to ultimately succumb to what are clearly complications of the operation. Finally, in order to discourage pre-terminal transfer of patients to other facilities in order to hide their anticipated deaths, report cards should include all patients who die within 30 days, regardless of cause or venue.

2.3.4 *Consumer choice.* Interestingly, despite the public accountability afforded by public report cards, there has been little objective evidence that this valuable information has redirected patients from high-mortality to low-mortality institutions. This was apparent in an early study by Vladeck and associates (Vladeck et al., 1988) following release of HCFA Medicare mortality data. The authors concluded that long-standing referral preferences, tradition, convenience and personal recommendation were more important than objective information. Similarly, Blendon et al. (1998) found that the recommendation of family and friends trumped objective data in choosing health care providers. Finlayson et al. (1999) determined that many patients preferred local care over demonstrably better care at regional referral centers, a strong geographic preference also shown by Shahian et al. (2000) for CABG surgery in Massachusetts. In Cleveland, an effort funded by local businesses to monitor and report quality of care did not demonstrate any measurable effect on consumer choice (Burton, 1999). Schneider and Epstein (1996, 1998) studied the responses of both cardiac surgery patients and cardiologists following release of the Pennsylvania cardiac surgery report card. They found that few patients were aware of the report card or knew their surgeon's rating prior to surgery. Few regarded it as important in their choice of a provider and few cardiologists felt that it had significant impact on their referral recommendations. Hannan et al. (1997c) found that only 38% of New York cardiologists thought that report cards had substantially impacted their referral patterns, despite regarding these report cards as readable and reasonably accurate. Furthermore, there was no shift in the market share of percentage of New York CABG patients who had surgery at high-mortality versus low-mortality hospitals after the introduction of report cards (Chassin, Hannan and DeBuono, 1996; Jha and Epstein, 2006).

It might be expected that payers, having easier access to outcome data, would be better able to direct patients to high-quality providers. However, in separate studies by Shahian and associates in Massachusetts (2000) and Erickson and colleagues in New York (2000), using completely different databases and statistical methodologies, both groups found ironically that managed care patients were less likely to have surgery at lower mortality hospitals. In general, promoting consumerism in health care has not been successful thus far. Consumer-driven health plans, which involve insurance arrangements that give employees greater choice among benefits and providers, but also expose them to greater financial risk, are the latest idea in health insurance. Early results indicate that beneficiaries in consumer-driven health plans have lower satisfaction, higher out-of-pocket costs and more missed health care than consumers in more comprehensive health insurance (Fronstin and Collins, 2005).

## 3. STATISTICAL ISSUES AND METHODS

### 3.1 Historical Approaches

Early attempts at measuring quality of care were based on tests of excess variation. These were soon replaced by methods that estimated quality measures and then attempted to identify poorly performing providers through tail probabilities. The first widely disseminated summary of variability in the quality of health care appeared in a 1973 *Science* article that examined medical and surgical rates across 193 hospital areas in New England (Wennberg and Gittelsohn, 1973). The authors quantified variation in dispensation of health care by the ratio of the maximum to the minimum rate, denoted the extremal quotient.

An obvious problem with the extremal quotient occurred in small areas where the minimum observed rate can often be zero. Other indices of variability emerged, including the coefficient of variation, CV (Chassin et al., 1986), and the systematic component of variation, SCV (McPherson et al., 1982). The former was used to quantify overall variation, while the SCV, defined as the difference between the total observed variation and the within-institution variation, was used as an estimate of interinstitution variability. However, these measures were shown to be sensitive to the number of institutions, the per-institution sample size and the underlying rate (Diehr et al., 1990).

The first HCFA report contained observed and expected mortality data for all acute-care nongovernmental hospitals. Hospitals with higher-than-expected mortality rates were classified as underperforming institutions. Let $Y_{ij}$ denote the outcome for the $j$th patient treated at the $i$th institution, let $\mathbf{x}_{ij}$ be a vector of patient-specific characteristics and let $n_i$ be the number of cases treated at institution $i$, where $i = 1, 2, \ldots, I$. Each hospital-specific mortality rate was calculated as the observed number of deaths at the institution divided by the number of cases, $\bar{y}_i = \frac{1}{n_i} \sum_j y_{ij}$. The expected mortality rate for a patient was modeled assuming

$$Y_{ij} \overset{ind}{\sim} \text{Bern}(p_{ij}),$$
$$\text{(1)}$$
$$\text{where } \text{logit}(p_{ij}) = \alpha_0 + \boldsymbol{\alpha}_1' \mathbf{x}_{ij}.$$

The covariates, $\mathbf{x}_{ij}$, were obtained from administrative claims data. The expected mortality rate in institution $i$ was calculated as

$$\text{(2)} \qquad \hat{y}_i = \frac{1}{n_{ij}} \sum_{j=1}^{n_{ij}} \text{logit}^{-1}(\hat{\alpha}_0 + \hat{\boldsymbol{\alpha}}_1' \mathbf{x}_{ij})$$

and compared to observed mortality using

$$\text{(3)} \qquad z_i = \frac{\bar{y}_i - \hat{y}_i}{\sqrt{\widehat{\text{Var}}(\bar{y}_i - \hat{y}_i)}}$$

where $\widehat{\text{Var}}(\bar{y}_i - \hat{y}_i)$ was approximated using a Taylor series expansion (Mood, Graybill and Boes, 1973). Hospitals with $z_i > 1.645$ were identified as *outliers* having higher than expected mortality.

In response to the earlier criticisms of the first mortality reports, several initiatives have been undertaken by HCFA (and now by CMS) to streamline in-depth data collection in order to better risk-adjust for case-mix differences. Nonetheless many report cards continue to use the observed data to calculate tail probabilities, for example, calculating $\frac{\bar{y}_i}{\bar{y}_i} \times \bar{y}$, a corresponding 95% CI, and classifying an institution as outlying if its 95% interval excludes $\bar{y}$.

## 3.2 Modern Approaches

Statistical researchers criticized the methodology utilized by HCFA/CMS on various methodological grounds (Thomas, Longford and Rolph, 1994; Goldstein and Spiegelhalter, 1996; Normand, Glickman and Gatsonis, 1997). The criticisms related to a lack of attention paid to the sampling variability due to large differences in the number of cases per hospital, ignoring the statistical dependence among outcomes within a hospital, failing to estimate inter- and intrahospital variance components, and utilization of a classification system that labels a predetermined number of hospitals as having quality problems when excess mortality could be due to random error. Table 1 demonstrates many of these problems using the Massachusetts data. The number of cases varies by an order of magnitude as do the observed mortality rates. If we assume an average expected mortality rate of 2.19%, then observing no mortalities at the new programs is not surprising. For example, in Hospital 5 with 26 cases the probability of no deaths is 0.56 and in Hospital 9 with 80 cases it is 0.17. On the other hand, there are 15 mortalities at Hospital 3. Ignoring case-mix, the probability of observing 15 mortalities for the 381 cases is 0.01, again assuming an underlying rate of 2.19%. Because patients are not randomized to hospitals, patient selection is a real issue—the last column of Table 1 illustrates this point. The expected mortality rate at the 13 institutions, estimated using $\hat{\mu}_i = \frac{1}{n_i} \sum_{j=1}^{n_i} E(Y_{ij} \mid x_{ij}, \hat{\boldsymbol{\beta}}_1, \hat{\mu}, \hat{\tau}^2)$ (Section 3.3), indicates that patients treated at Hospital 3 are relatively sicker, with an expected mortality rate of 2.94%, compared to those treated at Hospital 9 where the expected mortality rate is less than 1%.

To overcome the statistical shortcomings of the HCFA approach, researchers proposed the use of hierarchical models to describe hospital mortality,

$$\text{(4)} \qquad Y_{ij} \mid \beta_{0i} \overset{ind}{\sim} \text{Bern}(p_{ij}),$$

where $\text{logit}(p_{ij}) = \beta_{0i} + \boldsymbol{\beta}_1 \mathbf{x}_{ij}$,

$$\text{(5)} \qquad \beta_{0i} \overset{i.i.d.}{\sim} N(\mu, \tau^2).$$

In (5) $\tau^2$ represents between-hospital variation. The hierarchical model mimics the hypothesis that underlying quality leads to systematic differences among true hospital outcomes. If there are no between-hospital differences in mortality and $\mathbf{x}$ has been defined appropriately, then $\tau^2 = 0$ and $\beta_{01} = \beta_{02} = \cdots = \beta_{0I} = \beta_0$. While it is almost certain that $\tau^2 > 0$, the question is whether $\tau$ is small enough to ignore. An important feature of the hierarchical model relates to the multiple comparison issue. Multiplicity of parameter estimation is addressed by integrating all the parameters into a single model, for example, a common distribution for the $\beta_{0i}$'s. Regression to the mean is naturally accommodated because posterior estimates of the random intercepts, or of functions of the random intercepts, are "shrunk" toward the mean (Christiansen and Morris, 1997; Normand, Glickman and Gatsonis, 1997).

An implicit assumption in the model defined by (4)–(5) is that hospital mortality is independent of the number of patients treated at the hospital. While some researchers have shown volume to be related to mortality, the relationship between institutional volume and mortality is relatively weak in the case of CABG surgery (Peterson et al., 2004; Shroyer et al., 1996; Shahian and Normand, 2003).

### 3.3 Case-Mix Adjustments

The most controversial issue continually raised by institutions is the adequacy of risk adjustment. Because patients are not randomized to institutions, statistical adjustments are used to adjust for observed imbalances (Harrell, 2001). Adjustments are made through regression modeling although recent suggestions involve the use of propensity scores (Huang et al., 2005). The expected mortality rate at an institution is calculated as the number of expected deaths divided by the number of patients,

$$(6) \qquad \mu_i = \frac{1}{n_i} \sum_{j=1}^{n_i} E(Y_{ij} \mid x_{ij}, \boldsymbol{\beta}_1, \mu, \tau^2).$$

In addition to the type of analytical adjustment used, issues regarding inclusion of types of covariates are also important. For example, because administrative databases contain diagnoses upon hospital *discharge*, only those diagnostic codes that are thought to be present on admission are included in a risk model. For example, a discharge diagnosis of pneumonia, while predictive of mortality, may have arisen because of poor quality of care. Adjustment

TABLE 2
*Mean and adjusted odds ratios of 30-day mortality following isolated CABG surgery in adults, Massachusetts, 2002*

| Risk factor | Mean (%) | Odds ratio | 95% posterior interval |
|---|---|---|---|
| Years > 65[†] | 1.5 | 1.05 | 1.02, 1.07 |
| Male | 74.5 | 0.60 | 0.39, 0.96 |
| Renal failure | 7.3 | 2.39 | 1.32, 3.93 |
| Diabetes mellitus | 38.0 | 1.17 | 0.72, 1.76 |
| Hypertension | 77.0 | 2.91 | 1.35, 6.26 |
| Peripheral vascular disease | 18.0 | 1.73 | 1.05, 2.66 |
| Prior percutaneous coronary intervention | 18.6 | 0.87 | 0.48, 1.44 |
| Cardiogenic shock | 2.2 | 3.16 | 1.29, 6.45 |
| *Ejection fraction* | | | |
| ≥ 40 | 75.5 | 1.00 | — |
| < 30% or missing | 12.8 | 1.48 | 0.79, 2.44 |
| 30–39 | 11.7 | 1.33 | 0.68, 2.27 |
| *Myocardial infarction* | | | |
| No myocardial infarction | 51.1 | 1.00 | |
| Within 6 hours | 0.9 | 9.89 | 2.44, 26.63 |
| 7–24 hours | 1.8 | 3.72 | 1.15, 8.68 |
| 1–7 days | 20.7 | 1.10 | 0.57, 1.90 |
| 8–21 days | 5.7 | 1.45 | 0.56, 2.96 |
| > 21 days | 19.8 | 1.43 | 0.72, 2.54 |
| *Status of CABG* | | | |
| Elective | 34.0 | 1.00 | — |
| Urgent | 62.0 | 2.55 | 1.29, 4.81 |
| Emergent/salvage | 3.0 | 2.61 | 0.79, 6.44 |
| Pre-op intra-aortic balloon pump | 9.3 | 2.57 | 1.40, 4.37 |

Hierarchical model estimated using 4603 surgeries with 101 deaths.
[†]Represents the number of years over age 65 at time of surgery.

for such factors could "adjust out" the effect of interest. Table 2 displays the prevalence, adjusted odds ratios and corresponding 95% posterior interval estimates for the risk factors included in the Massachusetts CABG mortality model. Not surprisingly, cardiogenic shock and timing of myocardial infarction are the strongest predictors of 30-day mortality.

### 3.4 Identifying Underperforming Institutions

In addition to providing a standardized measure of outcome performance, virtually all report cards aim to identify institutions that are *outliers*. The key question is "what is an outlier?" The most common approach involves estimating an adjusted rate and identifying institutions in the tails of the distribution. Estimation of the adjusted rates, of course, involves specifying prior distributions for the hyper-parameters. The sensitivity of posterior inferences to the choice of the prior distributions is critical. Moreover, if outlying hospitals are present, their data can influence the distribution of between-hospital variance, $\tau^2$. Two alternative approaches to identifying outliers are to (1) specify the hyperparameters and determine which hospitals are "out-of-control" or (2) estimate the predicted number of mortalities at each hospital through cross-validation and compare the predicted number to the observed number.

### 3.4.1 Estimating the hyperparameters.

The model described by (4)–(5) assumes that the random effects are completely exchangeable, arising from a normal distribution with mean $\mu$ and variance $\tau^2$. The risk standardized mortality rate for each Massachusetts cardiac surgery program, for example, is

$$
\text{(7)} \quad
\begin{aligned}
&\text{Risk Standardized Mortality} \\
&= \left( \frac{\sum_j \frac{\exp(\beta_{0i} + \beta_1 \mathbf{x}_{ij})}{1 + \exp(\beta_{0i} + \beta_1 \mathbf{x}_{ij})}}{\sum_j \frac{\exp(\mu + \beta_1 \mathbf{x}_{ij})}{1 + \exp(\mu + \beta_1 \mathbf{x}_{ij})}} \right) \times 2.19
\end{aligned}
$$

and can be easily estimated using Markov chain Monte Carlo methods.

The ratio in the parentheses in (7) is the Bayesian version of the "O" to "E" ratio used in earlier versions of the HCFA reports. However, the "O" has been replaced by a shrinkage estimator that also adjusts for hospital case-mix. This ratio has a causal interpretation. Multiplying the ratio by the number of cases in the $i$th hospital and then subtracting the number of cases yields the number of excess mortalities (or if negative, the number of additional survivors) if the hospital's distribution of cases across risk categories had been what it was, but if its mortality rates across those risk categories were replaced by the state rates. The interested reader should see Draper and Gittoes (2004) and references therein where a counterfactual framework for estimators like that in (7) is discussed. The counterfactual distribution used in the Massachusetts report card is determined using the mortality risk observed in categories of patient types within the state and the prevalence of patient types observed within each hospital.

Figure 1 displays the risk standardized 30-day mortality rates for the 13 Massachusetts cardiac surgery programs and corresponding 95% intervals. The estimates are obtained using a vague proper conjugate prior distribution for $\tau^{-2}$, $\tau^{-2} \sim \text{Gamma}(0.001, 0.001)$, a vague but proper prior for $\mu$, $\mu \sim N(0, 1000)$, and similar independent vague normal priors for the components of $\boldsymbol{\beta}_1$. A burn-in of 5000 draws is used and a subsequent 3000 iterations for inference. The institution random effects are estimated by shrinking the risk-adjusted rates to $\mu$ where the amount of shrinkage is measured by the ratio of the within-institution variance to the total variance. While none of the intervals excludes the state unadjusted rate of 2.19%, the rate for Hospital 13 is clearly large with posterior mean 2.58 [median = 2.37].

The width of the interval estimate for Hospital 13 is surprisingly wide given the observed volume of 419 cases. Figure 2 displays the relationship between the numerator and denominator values simulated from the posterior distribution for four hospitals with varying volume: Hospital 5 with 26 cases, Hospital 7 with 718 cases, Hospital 8 with 149 cases, and Hospital 13 with 419 cases. The severity of the patient populations can be contrasted across the hospitals by examining the distribution of the draws on the $x$-axis and noting sicker populations are shifted to the right. The distribution of draws above the $x = y$ line in the graphs indicates increased binomial variability with the higher observed mortality rate at Hospital 13. The predicted probabilities at this institution have a skewed distribution.

*Sensitivity of posterior distribution to prior specification.* The sensitivity of posterior inferences to choice of prior distribution for $\tau^2$ is particularly important when comparing institutions (Gelman, 2002, 2006). The degree of sensitivity relates to the number of institutions or the sample size per institution. Spiegelhalter, Abrams and Myles (2004) have provided interpretations for plausible values for the standard deviation, $\tau$, in order to choose a prior. One interpretation involves specifying a plausible range for the ratio of the 97.5% odds of mortality to the 2.5% odds of mortality, say $a$, and then solving $\exp(3.92\tau) = a$ to determine a value for $\tau$. For example, when $\tau = 0.1$ the range in odds ratios is 1.48—the odds of dying at a "high"-mortality hospital relative to a "low"-mortality hospital—and this may be viewed as an acceptable range in variability for the random effects. A second method involves considering the absolute difference between a random pair of $\beta_{0i}$'s. The distribution of this difference, assuming normality for the random effects, is a half-Normal distribution with median $1.09\tau$. Thus, if a reasonable upper 95% point for $\tau$, $\tau_{0.95}$, can be specified, then $\tau \sim \text{half-Normal}((\tau_{0.95}/1.96)^2)$.

Table 3 illustrates the effects of the choice of prior specification of the variance component on posterior estimates of $\tau^2$ for the Massachusetts data. The

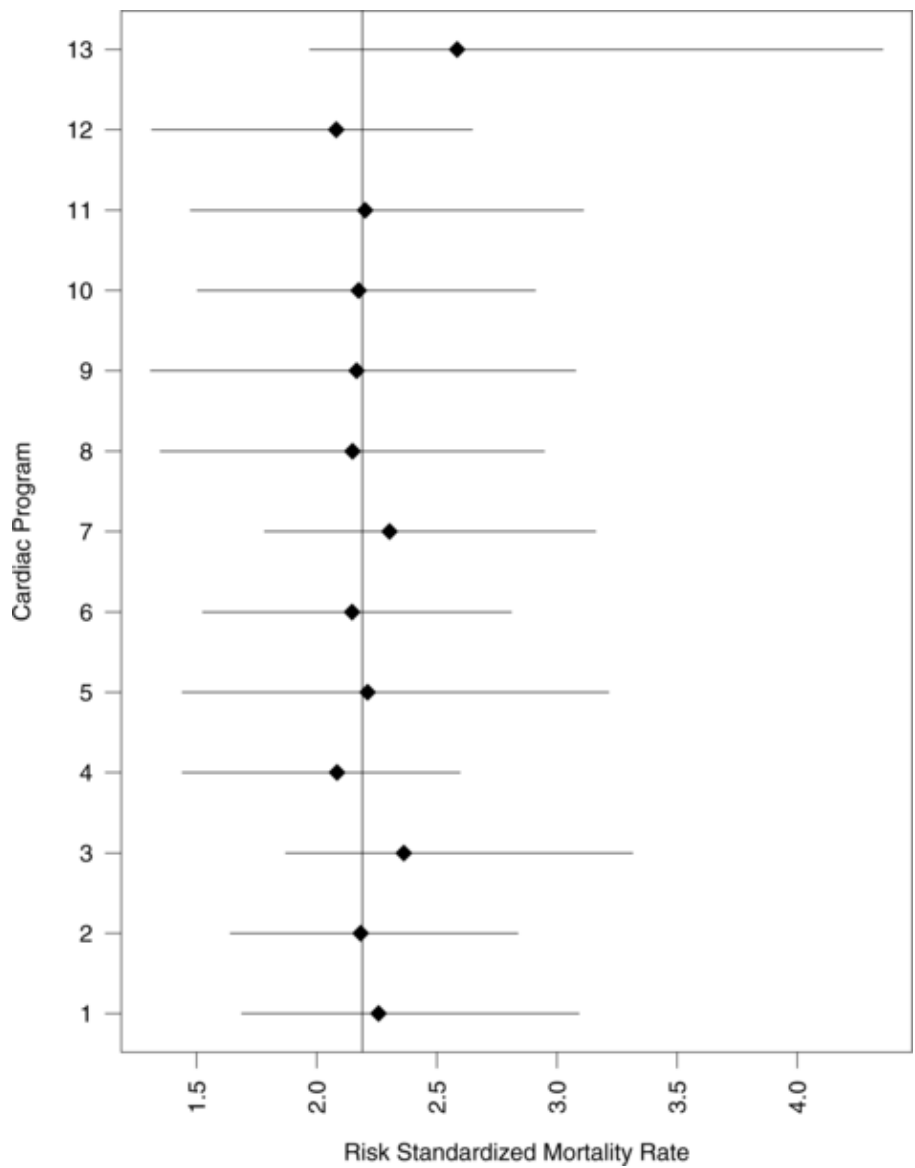

Fig. 1. *Risk standardized mortality rates (%) for 13 cardiac programs. Posterior mean and 95% posterior intervals.*

TABLE 3
*Sensitivity of posterior distribution of the hyperparameters to prior specification of*
*between-institution variance*

| Posterior summaries | | Prior distribution | | |
|---|---|---|---|---|
| | | $\tau^{-2} \sim$ **Gamma(0.001, 0.001)** | $\tau \sim$ **Unif(0, 1.5)** | $\tau \sim$ **half-Normal(0.26)** |
| $\tau^2$ | Mean | 0.04 | 0.08 | 0.09 |
| | Median | 0.02 | 0.04 | 0.05 |
| | 95% PI | (0.0007, 0.24) | (0.0007, 0.39) | (0.00009, 0.38) |
| $\mu$ | Mean | $-6.75$ | $-6.74$ | $-6.79$ |
| | Median | $-6.72$ | $-6.23$ | $-6.78$ |
| | 95% PI | $(-7.79, -5.88)$ | $(-7.97, -5.95)$ | $(-7.68, -5.75)$ |

PI = posterior interval.

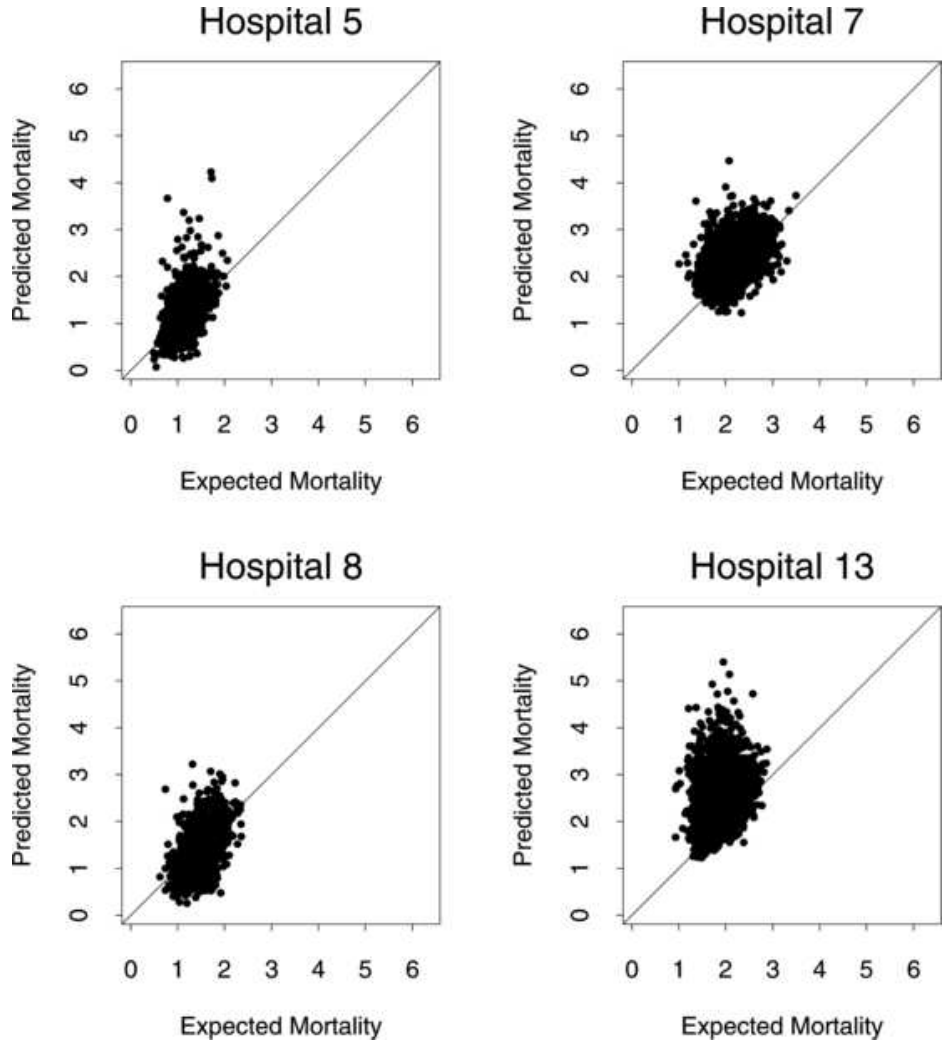

Fig. 2. *Expected versus predicted mortality rates (in %) for four cardiac surgery programs. Each plot displays the draws from the posterior distribution used for inference. The x-axis displays expected mortality rate [denominator in (7) divided by $n_i$] while the y-axis displays the shrinkage estimate [numerator in (7) divided by $n_i$] denoted "predicted" mortality in the figure. The solid line is the $x = y$ line.*

Gamma prior tends to place more weight on small values of $\tau^2$ than the other priors. A uniform prior for the standard deviation over the range 0 to 1.5 implies small values as equally likely as large values and the half-Normal uses $\tau_{0.95} = 1$. Not surprisingly, the posterior estimates of the between-program variance are smallest under the Gamma prior and largest under the half-Normal. Consequently the Gamma prior shrinks the program random effects toward the overall mean more than that with the other two prior distributions.

*Posterior predictive p-values.* A disadvantage of *estimating* the hyperparameters relates to estimation of $\tau^2$. If there is substantial between-institution variance, the posterior estimate of $\tau^2$ may actually mask outliers by accommodating a larger (than acceptable) estimate of between-hospital variance. A method to help detect this problem involves quantifying the discrepancy between the data and the model through replication (Gelman et al., 2004). For institutional profiling this idea is implemented by generating data sets with the same number of institutions, the same distribution of institution sample sizes and the same covariate distributions as those observed, and then comparing the observed number of mortalities at each institution to the posterior predictive distribution of the number of deaths,

$$\int_{\Omega} \int_{\bar{y}_i^r} I(\bar{y}_i^r \geq \bar{y}_i \mid \text{data})$$

(8)
$$\cdot f(\bar{y}_i^r \mid \Omega) f(\Omega \mid \text{data}) \, d\bar{y}_i^r \, d\Omega.$$

Here $\bar{y}_i^r$ denotes the mean mortality at institution $i$ in the replicated data set, $\Omega$ denotes the vector of model parameters $(\boldsymbol{\beta}, \mu, \tau^2)$ and $I(\bar{y}_i^r \geq \bar{y}_i \mid \text{data})$ is an indicator variable assuming a value of 1 if the replicated mean is larger than the observed mean. If the tail-area probability (denoted a posterior predictive $p$-value) is extreme, beyond 0.99 or below 0.01, and if the difference between the observed and replicated means is of practical significance, then this provides some evidence that the model for Hospital 13 is questionable. Column 2 in Table 4 lists the posterior predictive $p$-values when using a Gamma(0.001, 0.001) prior distribution for $\tau^{-2}$. Like the estimate of the risk standardized mortality rate, the $p$-values cast some suspicion on Hospital 13 ($p$-value = 0.03). The mean [95% PI] replicated mortality rate for Hospital 13 is 1.99% [0.7%, 3.82%] while the observed rate is 3.58%, a difference of practical importance.

3.4.2 *Specifying the hyperparameters.* Rather than estimating the hyperparameters, *acceptable* or *in-control* values of the hyperparameters could be pre-specified. This approach has the advantage of explicitly stating a standard and does not label a predetermined number of hospitals as having excess mortality. Posterior predictive $p$-values can be calculated in a way similar to that described in (8). Because the data contain more information about $\mu$ than $\tau$, we permit the data to estimate $\mu$. The key issue in this approach is how to select the values of $\tau$ that suggest "acceptable" variation. To determine the values, again, the guidance given by Spiegelhalter, Abrams and Myles (2004) is helpful.

Columns (3) and (4) in Table 4 list the posterior predictive $p$-values assuming the random effects arise from a Normal distribution with unknown mean and two different values for $\tau$. Specifying $\tau^2 = (0.10)^2$ implies that we are willing to accept 95% of the random effects to lie in a range of 1.5 in the odds ratio across the 13 cardiac programs. As in the other analyses, Hospital 13 appears on the boundary with a $p$-value of 0.02. The mean replicated observed mortality rate under this model is 1.96% compared to the observed rate of 3.58%. Using $\tau^2 = (0.01)^2$ indicates a willingness to tolerate virtually no between-hospital variation. Under this condition, the model effectively reduces to a logistic regression model with known intercept and no dependence on hospital. In this case (Column 4), the replicated rates should be close to the observed mean of 2.19% due to very little shrinkage. Hospital 12, which had an observed rate of 1.10%, is more than one full percentage point lower than the mean replicated rate ($p$-value = 0.01) when we specify a small value for $\tau^2$. If $\tau^2$ is made very large, then the model reduces to a logistic regression model with a fixed parameter for each hospital.

3.4.3 *Cross-validation.* Another method for identifying outlying hospitals is through cross-validation. In this approach, each hospital is dropped from the analysis, the parameters of the model, $\Omega$, are estimated, and then the mortality rate at the dropped institution is predicted by averaging over the posterior distribution. In a manner similar to the methods discussed in Sections 3.4.1 and 3.4.2, a posterior predictive $p$-value can be computed.

Column (5) of Table 4 presents the cross-validated $p$-values when systematically eliminating each hospital. Hospital 13 is again suspect with a posterior predictive $p$-value of 0.01. The remaining columns of Table 4 summarize the posterior estimates of the hyperparameters when excluding each hospital. The estimate of $\tau^2$ is substantially smaller when Hospital 13 is eliminated from the model, approximately 1/3 smaller than the estimate when this hospital is included in the model.

Other models are available to characterize outlyingness. For example, rather than assuming the institutional effects are completely exchangeable, partial exchangeability could be accommodated through a mixture model.

## 4. LOOKING AHEAD

The use of a single response to characterize an institution's quality of care, even when confined to care for a specific disease, is rather simplistic. In fact, under the leadership of the National Committee for Quality Assurance (NCQA), the Joint Commission on Accreditation of Healthcare Organizations (JCAHO), and various professional societies, consensus has emerged around core sets of process and outcomes measures for particular diseases and surgical procedures. NCQA, for example, sponsors and maintains the Health Plan Employer Data and Information Set (HEDIS) that consists of standardized performance measures and consumers' experiences for the purposes of comparing managed health

TABLE 4
*Analytical strategies for identifying poorly performing cardiac programs*

| Cardiac surgery program | Replication: posterior predictive $p$-values assuming $\beta_{0i} \sim N(\mu, \tau^2)$ | | | Cross-validation assuming $\beta_{0i} \sim N(\mu, \tau^2)$ | | |
|---|---|---|---|---|---|---|
| | $\tau^2$ unknown | $\tau^2 = (0.10)^2$ | $\tau^2 = (0.01)^2$ | $p$-value | $\mu_{-i}$ | $\tau^2_{-i}$ |
| (1) | (2) | (3) | (4) | (5) | (6) | (7) |
| 1 | 0.35 | 0.38 | 0.50 | 0.40 | −6.46 | $(0.174)^2$ $[(0.132)^2]$ |
| 2 | 0.53 | 0.48 | 0.45 | 0.49 | −6.73 | $(0.183)^2$ $[(0.143)^2]$ |
| 3 | 0.13 | 0.13 | 0.17 | 0.12 | −6.82 | $(0.182)^2$ $[(0.132)^2]$ |
| 4 | 0.26 | 0.22 | 0.13 | 0.22 | −7.19 | $(0.185)^2$ $[(0.145)^2]$ |
| 5 | 0.27 | 0.27 | 0.32 | 0.26 | −6.53 | $(0.158)^2$ $[(0.120)^2]$ |
| 6 | 0.38 | 0.37 | 0.27 | 0.30 | −6.95 | $(0.180)^2$ $[(0.146)^2]$ |
| 7 | 0.30 | 0.31 | 0.40 | 0.33 | −6.96 | $(0.187)^2$ $[(0.145)^2]$ |
| 8 | 0.36 | 0.34 | 0.25 | 0.35 | −6.77 | $(0.137)^2$ $[(0.103)^2]$ |
| 9 | 0.49 | 0.46 | 0.35 | 0.47 | −6.48 | $(0.149)^2$ $[(0.112)^2]$ |
| 10 | 0.46 | 0.42 | 0.35 | 0.45 | −6.60 | $(0.199)^2$ $[(0.152)^2]$ |
| 11 | 0.43 | 0.43 | 0.48 | 0.43 | −6.57 | $(0.184)^2$ $[(0.130)^2]$ |
| 12 | 0.20 | 0.16 | 0.01 | 0.17 | −6.69 | $(0.160)^2$ $[(0.123)^2]$ |
| 13 | 0.03 | 0.02 | 0.03 | 0.01 | −6.54 | $(0.127)^2$ $[(0.100)^2]$ |

All calculations assume $\mu \sim N(0, 1000)$. $\tau^{-2} \sim \text{Gamma}(0.001, 0.001)$ unless specified otherwise.

(2): Posterior probability observed mortality rate is more extreme than replicated mortality rate using all hospitals. Posterior mean [median] for $\mu = -6.75$ $[-6.72]$; $\tau^2 = 0.042$ $[0.016]$.

(3) and (4): Posterior probability observed mortality rate is more extreme than replicated rate using all hospitals and assuming an *in control* prior distribution. Posterior mean (SD) for $\mu$, Column 3: $-6.34$ $(0.392)$ and Column 4: $-6.76$ $(0.513)$.

(5): Posterior predictive probability that observed mortality rate is more extreme than the predicted mortality rate. Predictions use estimates based on all hospitals except $i$.

(6) and (7): Posterior mean average log-odds and mean [median] variance based on all hospitals except $i$.

care plans in the United States. The National Quality Forum has endorsed a set of 21 measures for assessing institutional CABG surgery quality (see a subset of measures in Table 5). The Society of Thoracic Surgeons has recently developed a multidimensional composite quality measure and rating system that utilizes structure, process and outcomes measures.

Interestingly, payors of health care have initiated pay-for-performance programs that offer bonus payment to hospitals that achieve high performance on the core sets (Galvin and Milstein, 2002; Milstein et al., 2000). Some health plans tier hospitals based on value similar to a tiered pharmacy benefit. Patients using hospitals classified in the high-value tier pay lower coinsurance or copayments at the point of care—often a 10% lower copayment (Steinbrook, 2004). The Medicare Modernization Act passed in 2003 established financial incentives for hospitals to provide CMS with data on quality indicators. The Hospital Quality Incentive Demonstration Project was launched in July 2003 to measure quality and pay incentives to participating hospitals that achieve "superior" levels of quality in five clinical areas

TABLE 5
*Examples of process-based measures of hospital quality for patients undergoing isolated CABG surgery*

| Measure | Description |
|---|---|
| Pre-operative beta-blockade | Percent of patients receiving beta-blockers within 24 hours preceding surgery |
| Use of internal mammary artery | Percent of patients receiving an internal mammary artery graft |
| Discharge Medications for In-Hospital Survivors | |
| Beta-blockade | Percent of patients discharged on beta blockers |
| Anti-platelet agent | Percent of patients discharged on anti-platelet therapy |
| Anti-lipid treatment | Percent of patients discharged on a statin or other pharmacologic lipid-lowering regimen |

(www.cms.hhs.gov). In a similar fashion, a consortium of organizations, including among others, CMS, the Joint Commission on Accreditation of Healthcare Organizations and the American Hospital Asso-ciation, initiated the Hospital Quality Alliance (`www.cms.hhs.gov/HospitalQualityInits/`). For fiscal years 2005 through 2007 eligible acute care hospitals reporting data to CMS receive an increase in the annual payment update from CMS for each of the DRG's under consideration (Jha et al., 2005). The data reported to CMS are standard clinical performance measures that have detailed specification, such as those listed in Table 5, and typically consist of a denominator that reflects the total number of patients eligible for a measure and a numerator that reflects the total number of eligible patients that received the therapy.

While a single measure may not be sufficient to describe the quality of care for a particular institution (Normand et al., 2007), the use of multiple measures can be challenging to consumers and to regulators (Epstein, 1998). Statistical issues regarding how to simultaneously model the multiple responses, how to derive a summary or composite measure, and how to select superiorly performing institutions arise.

To facilitate collection of performance measures, CMS has defined a set of Healthcare Common Procedure Coding System codes, termed G-codes, to supplement the usual claims data with clinical data. The goal is to use these new codes to define numerators and denominators for various performance measures. Availability of electronic health records may facilitate reporting of clinical data even further. Administrative data fashioned like California's inpatient discharge data supplemented with the new G-codes should lead to improved yet feasible databases.

### 4.1 Multiple Outcomes and Composite Measures

Let $Y_{ik}$ denote the number of patients receiving needed therapy $k$ at institution $i$ and let $n_{ik}$ denote the number of patients who should receive therapy $k$. For example, $n_{ik}$ may denote the number of patients undergoing CABG surgery who were discharged alive and $Y_{ik}$ is the number of these patients who were prescribed anti-lipid therapy.

In its Hospital Quality Incentive Demonstration Project, CMS calculates

$$\bar{y}_i = \frac{\sum_{k=1}^K y_{ik}}{\sum_{k=1}^K n_{ik}} \tag{9}$$

for each of $I$ hospitals and identifies hospitals falling into the 90th percentile of the empirical distribution of $\{\bar{y}_1, \bar{y}_2, \ldots, \bar{y}_I\}$. While such an approach is easily understood, it has several drawbacks. The optimal pooling algorithm depends on the type of measurement error associated with each measure, the correlation among the measures on the same individual, the scales of the measures and the missing data mechanism (see Horton and Fitzmaurice, 2004 for a review). A more recent proposal (Nolan and Berwick, 2006) advocates an "all-or-none" rule that constructs a binary response for each patient: a success is coded if the patient received all the care for which the patient was eligible; otherwise a failure is coded. This particular method addresses the within-patient issues but the variable number of eligible measures per patient is ignored and the other issues raised earlier remain.

The creation of composite measures to reflect performance is not new or unique to the assessment of health care. Lessons learned from education again are useful. The Stanford Achievement Test (SAT), first published in 1923, has been used to derive two composite scores, one for math and one for verbal ability, in order to assess student ability. The National Assessment of Education Progress (NAEP) is a U.S. congressionally mandated national survey to derive proficiencies scores to measure academic performance of U.S. students. Both the SAT and the NAEP use hierarchical models adapted from item response theory (IRT) to scale responses.

Using a similar approach, hospital composites could also be created. In the case of a collection of binary measures, the observed number of patients receiving needed therapy $k$ may be thought of as arising from a Rasch model (Rasch, 1960),

$$Y_{ik} \mid \beta, \theta_i \sim \text{Bin}(n_{ik}, p_{ik}),$$
$$\text{where } \text{logit}(p_{ik}) = \beta_{0k} - \beta\theta_i \tag{10}$$
$$\text{and } \theta_i \overset{i.i.d.}{\sim} N(0,1),$$

where $\beta_{0k}$ denotes the difficulty of measure $k$, $\theta_i$ denotes the underlying quality of the institution and $\beta$ is the precision of $\theta_i$. In (10) higher values of $\theta$ correspond to better quality of care, and would thus serve as the composite measure of hospital quality. If this model is commensurate with the data, then $\bar{y}_i$ is a minimally sufficient statistic for $\theta_i$ (Skrondal and Rabe-Hesketh, 2004). However, if the measures have different abilities to discriminate quality, then the following model is more reasonable:

$$Y_{ik} \mid \beta_k, \theta_i \sim \text{Bin}(n_{ik}, p_{ik}),$$
$$\text{where } \text{logit}(p_{ik}) = \beta_{0k} - \beta_{1k}\theta_i \tag{11}$$
$$\text{and } \theta_i \overset{i.i.d.}{\sim} N(0,1).$$

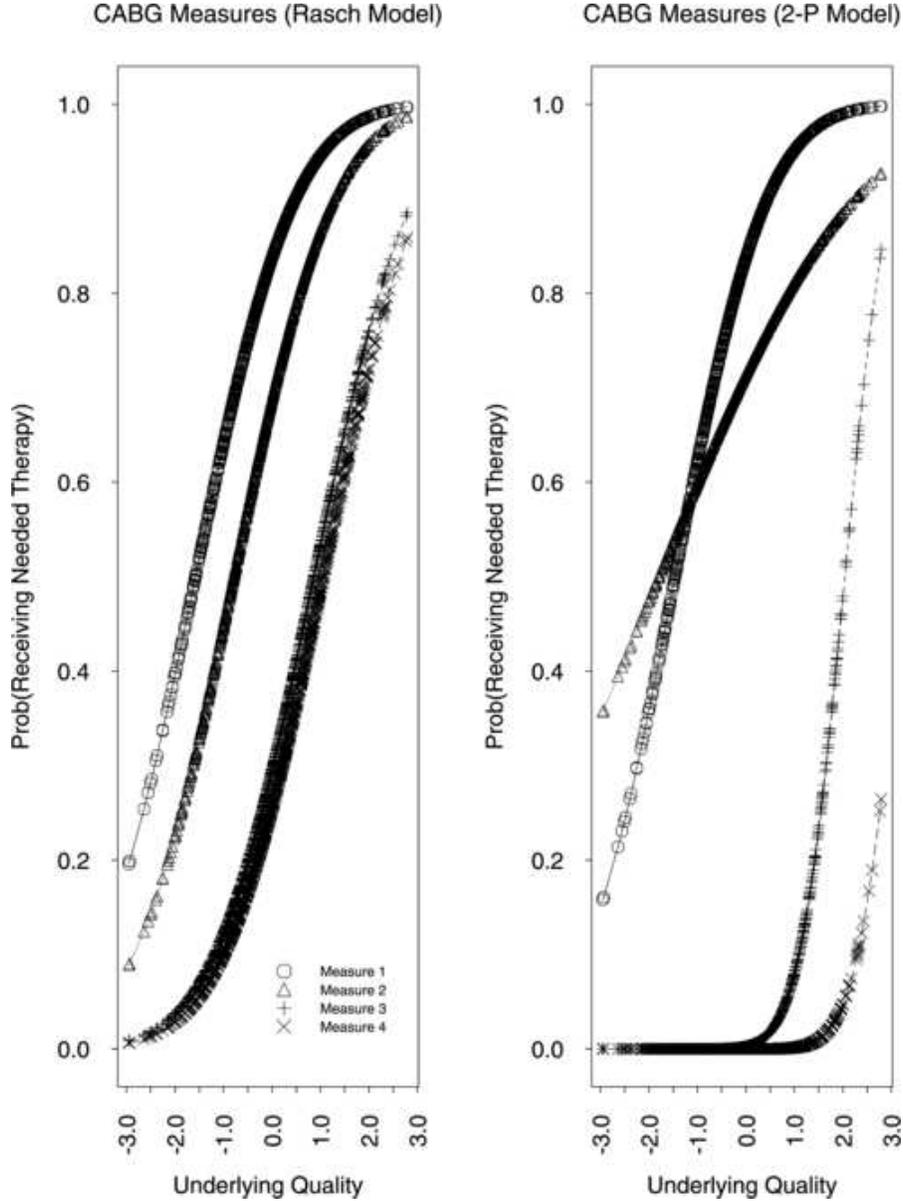

Fig. 3. *One-parameter (Rasch model) and two-parameter models for assessing hospital quality. The x-axis represents underlying hospital quality; the y-axis represents the probability of providing needed therapy. One-parameter model on left; two-parameter model on right.*

Here $\beta_{1k}$ measures how well the $k$th measure discriminates between hospitals with different qualities and is equivalent to a factor loading. The model in (11) is often referred to as a two-parameter logistic item response model while that in (10) is denoted a one-parameter logistic item response model. If (11) holds, then $\bar{y}_i$ is not sufficient for $\theta_i$ and financial incentives rewarded on the basis of the observed statistic could be distorted.

Figure 3 contrasts the relationship between the probability of providing needed therapy and insti-tutional quality assuming the model in (10) holds (left panel) with the relationship when the model in (11) holds (right panel). The figures were obtained through simulation and correspond to four hypothetical process-based measures. In the left panel, the four measures are equally discriminating of hospital quality, while the right panel illustrates that measures 3 and 4 are more discriminating of quality as demonstrated by the sharp changes from low to high probabilities of providing needed therapy. Landrum, Bronskill and Normand (2000) demon-strated these modeling issues when comparing hospital acute myocardial infarction quality.

As in the hierarchical model, the IRT models assume that the probability of receiving needed therapy is independent of the number of patients treated at the hospital. This assumption may be less believable in the context of process-based measures than in the context of mortality following CABG surgery. Furthermore, the IRT model assumes that different response measures within a hospital are conditionally independent given underlying hospital quality. There are two potential violations to this last assumption. First, patients contribute multiple measurements to the hospital composite. This issue can be easily addressed by adding another level to the IRT model, making the unit of observation the patient-measure and accounting for within-patient correlation. Even if the within-patient correlation is accommodated, there may be item-clustering (Scott and Ip, 2002; Bradlow, Wainer and Wang, 1999). This occurs if there are clusters of items or measures about a common stimulus used in assessing outcomes. While this is unlikely to arise in the context of process-based measures, it could occur when using patient surveys.

### 4.2 Multiple Mixed Outcomes

Efforts are underway to develop measures of hospital "efficiency" as reflected by costs, and to include these measures in pay-for-performance programs. Most efficiency measures assess technical efficiency defined as the cost of an episode of care using the least amount of resources. Because cost efficiency ignores information about health outcomes, some attempts have been made to examine cost and efficiency jointly. Specification of the joint distribution in the case of mixed outcomes has been the activity of recent methodological developments. If the measures are made on different scales but quantify the same underlying construct, then latent variable models (Sammel, Ryan and Legler, 1997) can be used to jointly model the observed outcomes. Latent variable models have also been extended to accommodate clustered outcomes (Dunson, 2000; Lee and Shi, 2001; Landrum, Normand and Rosenheck, 2003), although there has not been much practical experience. However, much less methodology and experience are available for modeling longitudinal mixed clustered outcomes (Daniels and Normand, 2006).

### 4.3 Concluding Remarks

While comparative profiling of health care institutions has been ongoing for more than a century, it has only been in the last decade and a half that statisticians have become actively involved. In this article, we reviewed the clinical considerations and implications of cardiac surgery profiling, and importantly, several methodological issues. It is intuitively obvious that there will be some between-institution variability. The conceptual issue relates to how much variability is acceptable, and how to quantify what is meant by overperforming and underperforming institutions. As profiling becomes increasingly linked to financial incentives, it is likely that the analytical methods for classifying institutions will be more closely scrutinized.

We concentrated on the analytical aspects of outcomes profiling, where design is an especially important consideration. Determination of power is particularly challenging in this area. Policy-makers will always be faced with low-volume providers and the difficulty of determining how best to characterize quality for providers with such small numbers. This problem can be particularly acute when comparing quality at the level of the individual physician (Landon et al., 2003).

Finally, we did not focus on the vexing issue of provider selection bias—unmeasured risk factors that confound case-mix with institutional quality. With more interest in causal inference, development of sensitivity analyses or instrumental variable analyses that can handle large numbers of "treatments" that characterize the profiling problem (13 in our cardiac surgery program example) will be critical.

### ACKNOWLEDGMENTS

The authors are indebted to Jennifer Grandfield, Ann Lovett, Treacy Silverstein, Robert Wolf and Katya Zelevinsky, all from the Department of Health Care Policy, Harvard Medical School, in Boston for data management, programming and technical support. Dr. Normand was supported in part by a contract from the Massachusetts Department of Public Health (620022A4PRE).